\documentclass[%
 reprint,
 amsmath,amssymb,
 aps,
]{revtex4-2}

\usepackage{hyperref}
\usepackage{placeins}
\usepackage{graphicx}
\usepackage{dcolumn}
\usepackage{bm}

\begin{document}

\preprint{APS/123-QED}

\title{Beam energy dependence of transverse momentum distribution and elliptic flow in Au-Au collisions using HYDJET++ model\\}

\author{Satya Ranjan Nayak$^{1}$}
 \email{satyanayak@bhu.ac.in}
\author{Saraswati Pandey$^{1,3}$}
 \email{saraswati.pandey13@bhu.ac.in}
\author{B. K. Singh$^{1,2}$}%
 \email{bksingh@bhu.ac.in}
 \email{director@iiitdmj.ac.in}
\affiliation{$^{1}$Department of Physics, Institute of Science,\\ Banaras Hindu University (BHU), Varanasi, 221005, INDIA. \\
$^{2}$Discipline of Natural Sciences, PDPM Indian Institute of Information Technology Design \& Manufacturing, Jabalpur-482005, India\\
$^{3}$Department of Physics, University of Virginia, Charlottesville, VA 22904, USA}
\date{\today}

\date{\today}
\begin{abstract}

In this work, we present the particle ratios, transverse momentum spectra, and elliptic flow ($v_2$) of $\pi^\pm,k^\pm,$  $p$, and $\bar{p}$ in Au-Au collisions at $\sqrt{s_{NN}}=$ 62.4, 39.0, 27.0, 19.6 and 11.5 GeV using HYDJET++ model. The particle ratios match the experimental data that validates the Cleymans-Reidlich parameterization of freeze-out parameters at lower beam energies under the HYDJET++ framework. The lower collision energies produce a system of high baryon chemical potential ($\mu_B$) and have a lower inelastic cross section. The interplay between these effects affects the overall shape of the $p_T$ spectra. The HYDJET++ model calculations for $p_T$ spectra agree well with the available experimental data. The invariant yield ratio of central and peripheral collisions is independent of beam energy. The elliptic flow is calculated based on the scaling between initial and final azimuthal spatial anisotropy ($k$). This interpretation of $v_2$ successfully describes the experimental data for all the collision energies studied in this work. The positive correlation of $k$ with beam energy leads to a small $v_2$ at lower collision energies. The hadrons containing strange quarks tend to have smaller values of $k$ than the non-strange hadrons.

\end{abstract}

\maketitle


\section{\label{sec:level1}Introduction\protect\\ }
Heavy Ion collision experiments at RHIC and LHC have a common objective to study an exotic state of matter called Quark Gluon Plasma(QGP) formed at extreme temperature and density, which is strong enough to de-confine the quarks and gluons \cite{Collins:1974ky,Cabibbo:1975ig,Chapline:1976gy}. One of the plausible approaches to study such a de-confined medium is to vary the collision energy and study its signals \cite{singh1993signals} like elliptic flow \cite{Wiedemann:1997cr}, jet quenching\cite{PHENIX:2001hpc}, strangeness abundance \cite{Rafelski:1982pu,Rafelski:1982ii}, $j/\psi$ suppression \cite{Wong:1997rm}, etc. Such an approach can be used for scanning the phase diagram of QCD as both temperature and chemical potential can be changed by varying collision energy \cite{Becattini:2005xt,Andronic:2005yp,Cleymans:1999st}. Another advantage of using such a method is the possibility of finding the QCD critical point, which marks the end of the first-order phase transition \cite{Fodor:2004nz,Gavai:2008zr}. The critical point can be identified by enhanced fluctuations in conserved quantities like baryon number, charge, strangeness, etc \cite{STAR:2017tfy,STAR:2019ans,Mohanty:2009vb}. 

An experimental approach to study the collision systems at wide range of beam energies was made in the RHIC Beam Energy Scan(BES) program \cite{STAR:2013ayu,STAR:2017sal,STAR:2009sxc} which explored thirteen beam energies from the top RHIC energy of $\sqrt{s_{NN}}=$ 200 GeV to the lowest energy of 3 GeV. A series of experiments under the BES program tried to find the phase boundary and critical point by looking for the signals of QGP at different collision energies. The BES program generated a large volume of experimental data for different important observables like $p_T$ spectra, $v_2$, higher moments, etc, for different collision energies. The highest energy of the BES program (i.e. 200 GeV) has been studied widely, while more phenomenological studies need to be performed for these smaller systems to study their behavior and mechanism of particle production. Furthermore, a qualitative difference between experimental data and theoretical model calculations or a deviation from the trend of model parameters at a particular beam energy could suggest a change in the particle production mechanism. Hence, the beam energy dependence of different signals and the model parameters need to be studied using a suitable theoretical model.

 Some theoretical attempts were made before to predict the beam energy dependence of signals like $v_2$ using models like AMPT, UrQMD, etc \cite{Solanki:2012ne,Nasim:2010hw}. A hydrodynamical treatment can be used to get a new insight into this problem. The theoretical model we chose to employ is the HYDJET++ model. It gives the option to study the collision systems as a combination of soft QCD processes (HYDrodynamics) and hard QCD processes (JETs). Our main goal is to study the bulk properties of the QGP medium produced in lower collision energies such as $p_T$ spectra, particle multiplicities, and elliptic flow, which are best described by the Monte Carlo event generators. The HYDJET++ model has been successful in reproducing the experimental data at LHC energies and the top RHIC energy \cite{Lokhtin:2009be,Lokhtin:2008xi,Singh:2023bzm}. One of the main distinctions between the top RHIC energy, the LHC energies, and these lower energies is the large values of the chemical potential of these smaller systems. The model needs to be tested at lower collision energies to check the validity and implications of such treatment.

The paper is organized as follows. In section II, we have discussed the HYDJET++ framework. Section III is divided into three sub-sections. In sub-section A, we have discussed the particle ratios and freeze-out parameters. In sub-section B, we have shown the $p_T$ spectra at different collision energies and discussed the contribution from hard and soft processes. In sub-section C, we have shown the elliptic flow $v_2$ as a function of $p_T$ at different collision energies. Finally, in section IV, we have summarised our work.

\section{HYDJET++ MODEL}
HYDJET++ is a Monte Carlo Heavy Ion collision event generator that produces particles as a combination of hard and soft processes \cite{Lokhtin:2008xi,Lokhtin:2010zz}. The soft part does a detailed treatment of thermal particle production, collective flow, etc. While the hard part deals with jet production, collisional and radiative energy loss, nuclear shadowing, etc.

The hard part of the model is generated by PYQUEN. The initial patron spectra and jet production vertices from binary N-N sub-collisions are generated by PYTHIA 6 \cite{Sjostrand:2006za}. The jets produced in  Au-Au collisions are a combination of jets produced from these binary sub-collisions. PYQUEN calculates the rescattering path of partons in the dense QGP medium and the radiative and collisional energy loss. The final hadronization is done according to the Lund string model. The nuclear shadowing for each impact parameter is done according to the Glauber-Gribov theory \cite{Tywoniuk:2007xy}.

The soft part of the HYDJET++ is generated on chemical and thermal freeze-out hypersurfaces with $T_{th} $ and $T_{ch} $ as the input parameters. A detailed physics framework can be found in the corresponding papers \cite{Amelin:2006qe,Amelin:2007ic}. The momentum distribution of hadrons is given by the equilibrium  distribution function:
\begin{eqnarray}
f_{i}^{eq}(p^{*0},T,\mu_i,\gamma_s)=\frac{g_i}{\gamma_{s}^{-n_{i}^s}exp([p^{*0}-\mu_i]/T)\pm1}
\label{eq:1}
\end{eqnarray}

Where $p^{*0}$ is the hadron energy in the fluid rest frame, $g_i = 2J_i+1$ is the spin degeneracy, $\gamma^s$ is the strangeness suppression factor, and $\pm$ takes account of the quantum statics of a fermion or boson. 
The particle density at a given $T_{ch} $ and $\mu$ is given by
\begin{eqnarray}
\rho_i^{eq}(T,\mu_i)=\frac{g_i}{2\pi^2}m_i^2T\sum_{k=1}^{\infty}\frac{(\mp)^{k+1}}{k}exp(\frac{k\mu_i}{T})K_2(\frac{km_i}{T})
\label{eq:2}
\end{eqnarray}
where $m_i$ is the particle mass and $K_2$ is the modified Bessel's function.
Chemical potential $\mu_i$ of a particle depends on baryon number B, strangeness S, electric charge(Isospin) Q, charm C, etc. However, for this work, we have considered only baryon number, strangeness, and i.e. $\mu_i=B_i\mu_B+S_i\mu_s+Q_i\mu_Q $. The baryonic chemical potential and chemical freeze-out temperature for different beam energies are chosen according to the following equation \cite{Cleymans:1999st}.
\begin{eqnarray*}
T(\mu_B)=a-b\mu_B^2-c\mu_B^4,\\
\end{eqnarray*}
\begin{eqnarray}
\mu_B(\sqrt{s_{NN}})=\frac{d}{1+e\sqrt{s_{NN}}}
\label{eq:3}
\end{eqnarray}
where, a=0.166$\pm$0.002 GeV, b=0.139$\pm$0.016 Ge$V^-1$,\\c=0.053$\pm$0.021 Ge$V^-3$,d=1.308$\pm$0.028 GeV, e=0.273$\pm$0.008 Ge$V^-1$.\\
The values of $\mu_s$ are chosen to get the best match with the experimentally observed $k^-/k^+$, $k^+/\pi^+$ and  $k^-/\pi^-$ ratios and $p_T$ spectra simultaneously. The value of isospin chemical potential is kept constant for all beam energies ($\mu_Q=-0.001$ GeV). The particle numbers are fixed after the chemical freeze-out. However, there are some corrections in the number of produced particles due to the decay of short-lived particles. To estimate the particle number densities at thermal freeze-out, the particle ratios at chemical freeze-out are assumed to be the same as thermal freeze-out. The effective pion chemical potential at thermal freeze-out is defined as follows,
\begin{eqnarray}
\frac{\rho_i^{eq}(T^{ch},\mu_i)}{\rho_{\pi}^{eq}(T^{ch},\mu_i^{ch})}=\frac{\rho_i^{eq}(T^{th},\mu_i^{th})}{\rho_{\pi}^{eq}(T^{th},\mu_\pi^{eff})}
\label{eq:4}
\end{eqnarray}
where, $\mu_{\pi^+}^{eff}$ is the effective chemical potential of $\pi^+$ at thermal freeze-out. The values of $T_{th}$ and $\mu_{\pi^+}^{eff}$ are chosen to get the best match with $p_T$ distribution measured in RHIC BES \cite{STAR:2017sal}.
Assuming Boltzmann approximation in eq. ~(\ref{eq:2}) for particles heavier than pions \cite{Yen:1997rv}, from eq. ~(\ref{eq:2}) and eq. ~(\ref{eq:4}) the chemical potential of $i^{th}$ hadron species at thermal freeze-out:
\begin{eqnarray}
\mu_{i}^{th}=\ln{(\frac{\rho_i^{eq}(T^{ch},\mu^{ch}_i)}{\rho_{i}^{eq}(T^{th},\mu_i=0)}\frac{\rho_{\pi}^{eq}(T^{th},\mu_i^{th})}{\rho_{\pi}^{eq}(T^{th},\mu_\pi^{eff})})}
\label{eq:5}
\end{eqnarray}

The longitudinal motion is dominant in the relativistic heavy ion collisions. Hence, the fluid flow four-velocity $\{u^0(x),\vec{u}(x)\}$ is parameterized in terms of longitudinal (z) and transverse ($r_\perp$) flow rapidities

\begin{eqnarray}
\eta^u(x)=\frac{1}{2}\ln{\frac{1+v_z(x)}{1-v_z(x)}}
\label{eq:6}
\end{eqnarray}
\begin{eqnarray}
\rho^u(x)=\frac{1}{2}\ln{\frac{1+v_\perp(x)cosh\eta^{u}(x)}{1-v_\perp(x)cosh\eta^{u}(x)}}
\label{eq:7}
\end{eqnarray}
Where $v_\perp$ is the magnitude of the transverse component of flow three-velocity $\vec{v}=\{v_\perp,v_z\}=\{v_\perp cos\phi^u,v_\perp sin\phi^u,v_z\}$, i.e,
\begin{eqnarray*}
u^\mu(x)=\{cosh\rho^ucosh\eta^u,sinh\rho^ucos\phi^u,\\sinh\rho^usin\phi^u,cosh\rho^usinh\eta^u\}
\label{eq:8a}
\end{eqnarray*}
\begin{eqnarray}
=\{(1+u_{\perp}^2)^{1/2}cosh\eta^u,u_\perp,(1+u_{\perp}^2)^{1/2}sinh\eta^u\}
\label{eq:8}
\end{eqnarray}
$u_\perp=\gamma v_{\perp}=cosh\eta^u\gamma_{\perp}v_{\perp}$, $\gamma_{\perp}=cosh\rho^u$.
Due to the presence of non-zero azimuthal anisotropy, the azimuthal angle of the fluid velocity vector is not identical to the spatial azimuthal angle, i.e.
\begin{eqnarray*}
u^\mu(x)=\{\gamma^\phi cosh\rho^{'u} cosh\eta^u,\sqrt{1+\delta(b)}sinh\rho^{'u}cos\phi,\\
\sqrt{1-\delta(b)}sinh\rho{'u}sin\phi,cosh\rho^{'u}sinh\eta^u\}
\label{eq:9a}
\end{eqnarray*}
Where,\\
\begin{eqnarray*}
\gamma^\phi=\sqrt{1+\delta(b) tanh^2\rho^{'u} cos 2\phi}\\
tan\phi^u=\sqrt{\frac{1-\delta(b)}{1+\delta(b)}} tan\phi
\label{eq:9b}
\end{eqnarray*}
The parameter $\delta(b)$ is the azimuthal momentum anisotropy, a more detailed description of $\delta(b)$ is given later in the text. The transverse flow rapidity $\rho^u$ is related to $\rho^{'u}$ as follows:
\begin{eqnarray*}
u_\perp=sinh \rho^u=\sqrt{1+\delta(b)cos2\phi} sinh\rho^{'u}
\label{eq:9c}
\end{eqnarray*}
We have used a simple rapidity linear profile for $\rho^{'u}$ i.e
\begin{eqnarray}
\rho_u=\frac{r}{R_f(b)}\rho_u^{max}(b=0)
\label{eq:9}
\end{eqnarray}
where, r is the radial co-ordinate,  $\rho_u^{max}(b=0)$ is the maximum transverse flow rapidity in central collisions. $R_f(b)$ is the mean-square radius of the hadron emission region.

In non-central collisions, the shape of the emission region in the x-y plane can be approximated by an ellipse with semi-major axis $R_y$ and semi-minor axis $R_x$ \cite{Amelin:2007ic}. The magnitude of the $R_{f}=\sqrt{(R_{x}^2 + R_{y}^2)/2 }$. $R_x(b) = R_{f}\sqrt{1-\epsilon(b)}$ and $R_y(b) = R_{f}\sqrt{1+\epsilon(b)}$, where $\epsilon(b)=\frac{R_{y}^2-R_{x}^2}{R_{y}^2+R_{x}^2}$. From the ellipse equation follows the dependence of the transverse radius of the hadron emission region on $\phi$
\begin{eqnarray}
R(b,\phi)=R_f(b)\frac{\sqrt{1-\epsilon^2(b)}}{\sqrt{1+\epsilon(b)cos 2\phi}}
\label{eq:10}
\end{eqnarray}
Here, $\epsilon$ is the spatial anisotropy. 

The effective volume of the hadron emission region for hypersurface of proper time $\tau$ is calculated as follows:\\

\[V_{eff}=\tau \int_{0}^{2\pi} \,d\phi \int_{0}^{R(b,\phi)} (n_\mu n^\mu)\,rdr \int_{\eta_{min}}^{\eta_{max}}f(\eta) \,d\eta \]
 Where $(n_\mu n^\mu)=cosh\rho_u\sqrt{1+\delta(b)tanh^2\rho_ucos2\phi}$ and $f(\eta)$ is the longitudinal rapidity profile (assumed uniform for this work) and $\eta_{max}$ is the maximal longitudinal flow rapidity.

Elliptic flow is a result of pressure gradient in the overlap region. The expansion would be stronger along the reaction plane where the pressure gradient is larger. Due to this expansion, the initial spatial anisotropy keeps changing \cite{Kolb:1999it}. Since the model does not trace the evolution of spatial anisotropy, we have only used the spatial anisotropy at thermal freeze-out. The initial spatial anisotropy depends on the impact parameter and nuclear geometry. Hence, we have assumed a simple scaling between the eccentricity of the elliptical overlap region($\epsilon_0=b/2R_A$) and the spatial anisotropy at thermal freeze-out, i.e.
\begin{eqnarray}
\epsilon=k\epsilon_0
\label{eq:11}
\end{eqnarray}
The parameter ``$k$" is a scaling factor that indicates the change in azimuthal spatial anisotropy during evolution. It is a common practice to treat the parameters $\epsilon$ and $\delta$ as free parameters, but in such a treatment, the parameters lose all relation with the initial collision geometry. The relation can be established through the scaling factor ``$k$". The scaling factor relates the eccentricity of the overlap region $\epsilon_0$ with the azimuthal spatial anisotropy of the fireball at freeze-out $\epsilon$, which then determines the momentum anisotropy $\delta$. Hence, the scaling factor ``$k$" determines both the spatial and momentum anisotropy parameters and the elliptic flow of final state hadrons.  An empirical relation between the parameters estimated by the hydro-dynamical approach in the corresponding paper \cite{Wiedemann:1997cr}:
\begin{eqnarray}
v_2 \propto \frac{2(\delta - \epsilon)}{(1-\delta^2)(1-\epsilon^2)}
\label{eq:12}
\end{eqnarray}

Using the fact that $v_2\propto\epsilon_0$ and $\epsilon \propto \epsilon_0$, the relation between $\epsilon$ and $\delta$ can be calculated using eq. ~(\ref{eq:12}),
\begin{eqnarray}
\delta=\frac{\sqrt{1+4B(\epsilon+B)}-1}{2B},B=C(1-\epsilon^2)\epsilon
\label{eq:13}
\end{eqnarray}
The conversion of spatial anisotropy to momentum anisotropy, which is controlled by the parameter ``C", is assumed to be constant across all collision energies. In this work, we have fixed the value of $\mathrm{C=2} $, which was used at 200 GeV in the corresponding paper \cite{Lokhtin:2008xi}. The value of the scaling factor ``$k$" is chosen to get the best match with the experimental data.
\begin{figure}[h]
\includegraphics[width=.52\textwidth]{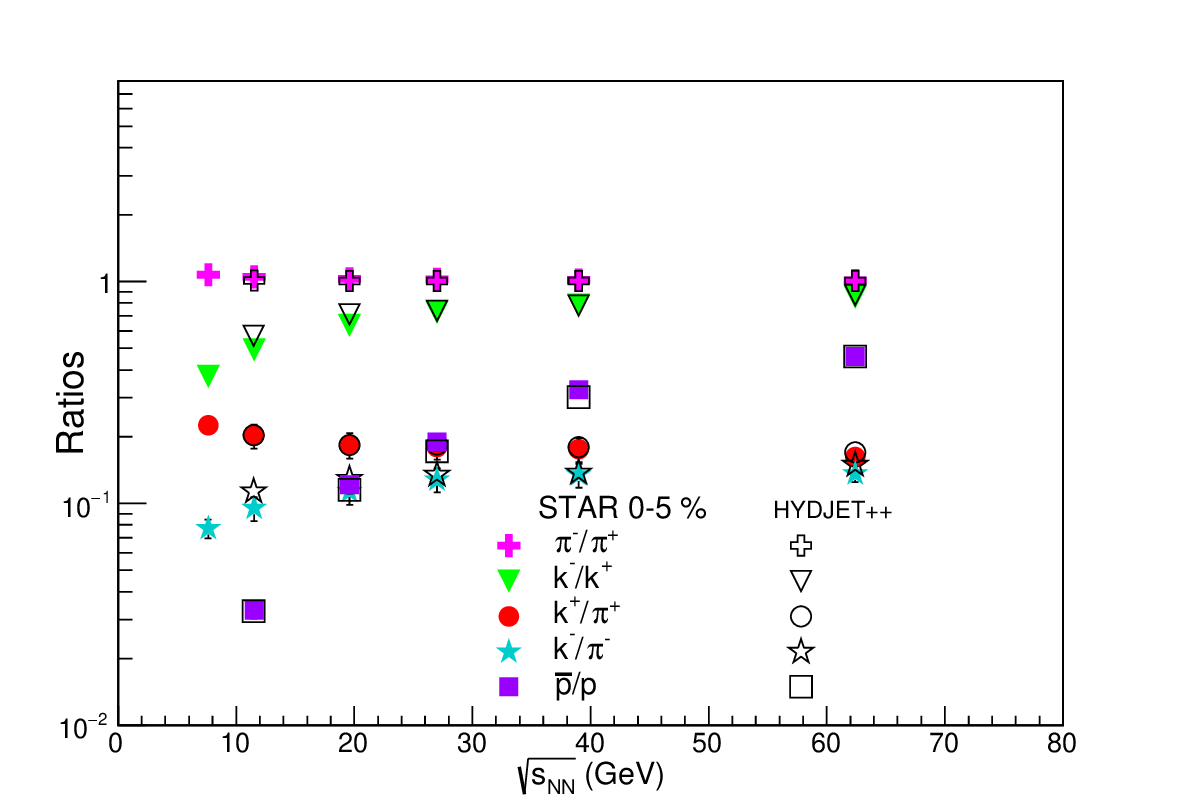}
\caption{\label{fig:two}Particle ratios at different beam energies.}
\end{figure}
\begin{figure}[h]
\includegraphics[width=.52\textwidth]{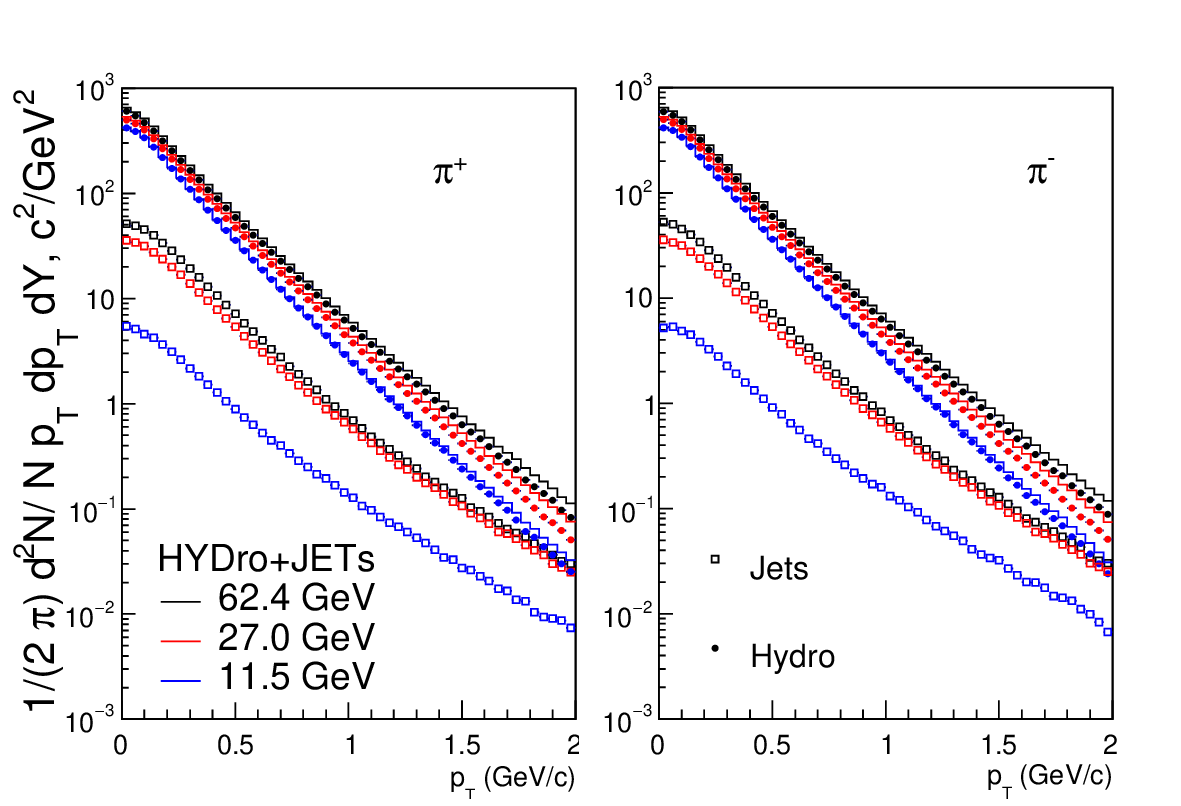}
\caption{\label{fig:1}Contributions from hydro and jet for $\pi^\pm$.}
\end{figure}
\begin{figure}[h]
\includegraphics[width=.52\textwidth]{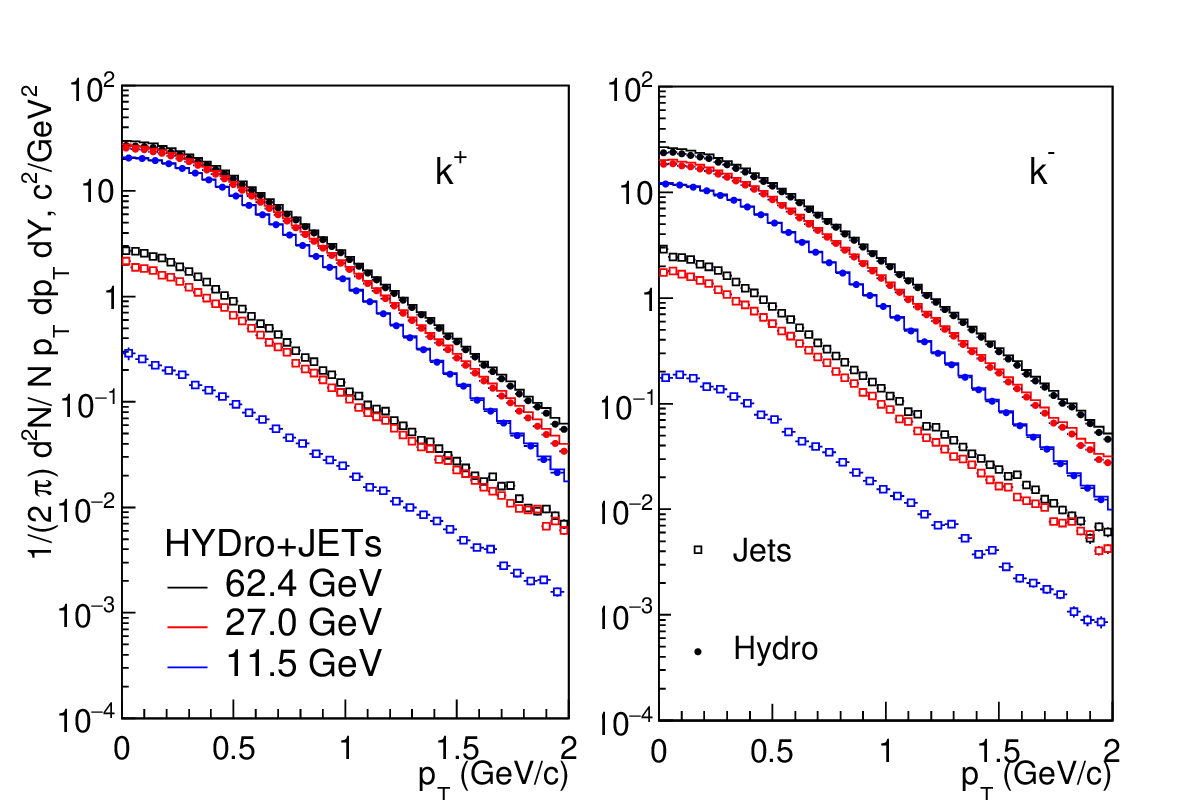}
\caption{\label{fig:2}Contributions from hydro and jet for $k^\pm$.}
\end{figure}
\begin{figure}[h]
\includegraphics[width=.52\textwidth]{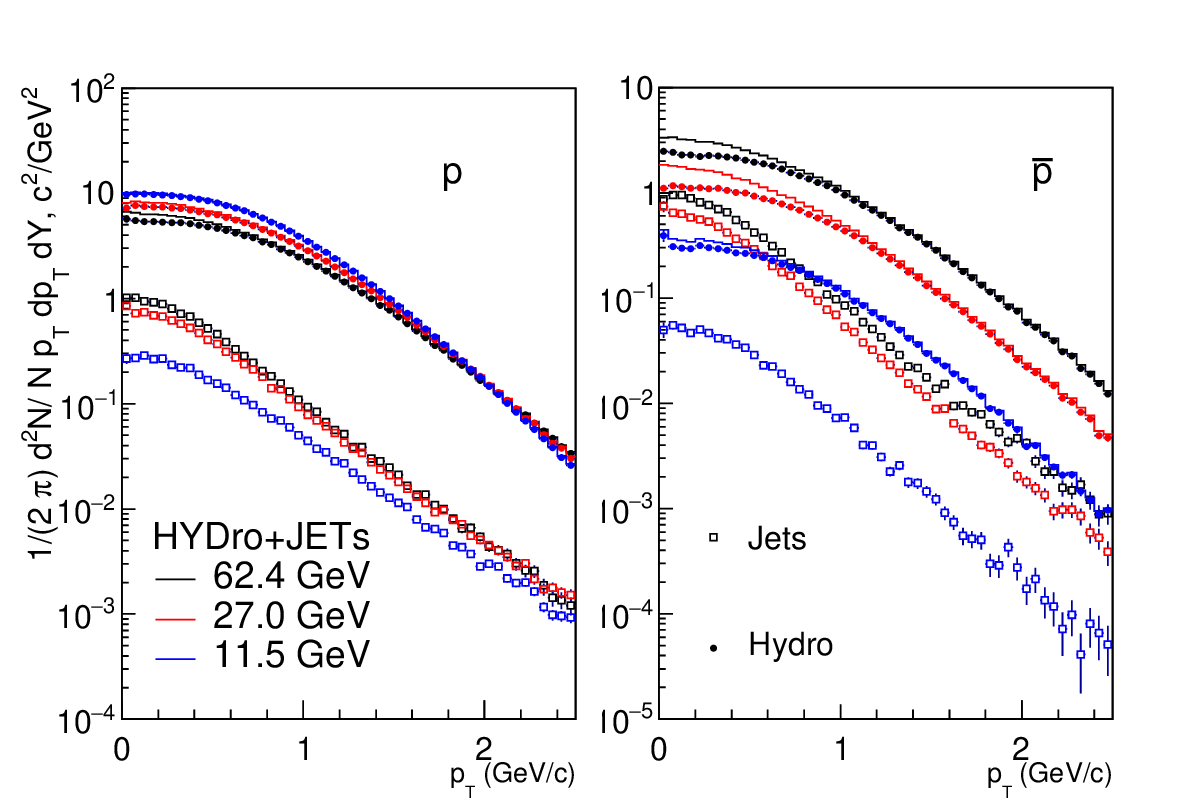}
\caption{\label{fig:3}Contributions from hydro and jet for $p$ and $\bar{p}$.}
\end{figure}

\begin{figure*}
\includegraphics[width=.72\textwidth]{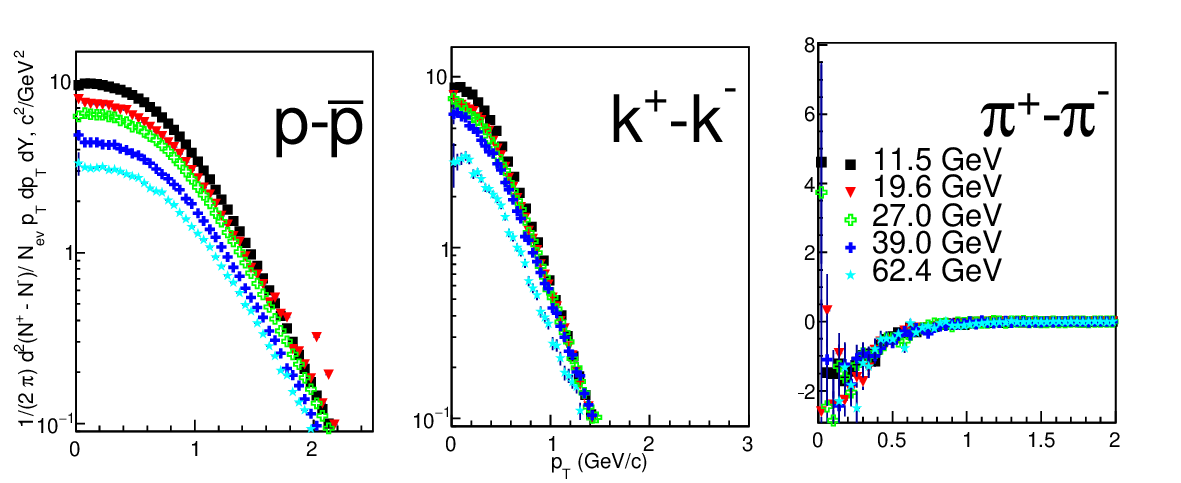}
\caption{\label{fig:4}Difference in invariant yield of particles and antiparticles as a function of beam energy.}
\end{figure*}

Besides the above-synchronized formalism, the HYDJET++ model also has certain limitations. HYDJET++ model gives good results for the central and semi-central collision events, but it fails to give convincing results for the peripheral collisions. The model can only be used for symmetric collisions (i.e., both the colliding nuclei should be of the same element). It works reliably for relatively heavier nuclei with nuclear mass A$>$40. The model best describes the experimental data in the mid-rapidity region. Another limitation of the HYDJET++ model is that it should be used for $\sqrt{s_{NN}}>10 $ GeV. For the limitations described above, we have chosen to limit our results up to 11.5 $\mathrm{GeV}$ and presented all our results in the mid-rapidity region.
\section{RESULTS AND DISCUSSIONS}
We have simulated Au-Au collisions at center-of-mass energies 62.4 GeV, 39.0 GeV, 27.0 GeV, 19.6 GeV, and 11.5 GeV using the HYDJET++ event generator and compared our results with the experimental data from STAR and PHENIX detectors. The kinematic range of our results is kept the same as the experimental data. The rapidity range for particle ratios and $p_T$ spectra is $|y|\le 0.1$, and for $v_2$ rapidity range is $|y|\le$ 0.35 for PHENIX data and $|y|\le$ 1.0 for STAR data. The exact number
of events can be found in Table~\ref{tab:table1}.

\begin{table}[h]
\caption{\label{tab:table1}%
The number of events for each centrality class.
}
\begin{ruledtabular}
\begin{tabular}{ccccc}
\textrm{$\sqrt{s_{NN}}$ (GeV)}&
\textrm{0-5\%}&
\textrm{10-20\%}&
\textrm{30-40\%}&
\textrm{50-60\%}\\
\colrule
62.4 & 1.5x$10^5$ & 2.0x$10^5$ & 2.5x$10^5$ & 2.0x$10^5$\\
39.0 & 2.0x$10^5$ & 2.5x$10^5$ & 3.0x$10^5$ & 5.0x$10^5$\\
27.0 & 2.0x$10^5$ & 3.5x$10^5$ & 5.0x$10^5$ & 7.0x$10^5$\\
19.6 & 3.0x$10^5$ & 4.5x$10^5$ & 5.0x$10^5$ & 7.0x$10^5$\\
11.5 & 5.0x$10^5$ & 7.0x$10^5$ & 8.0x$10^5$ & 1.0x$10^6$\\
\end{tabular}
\end{ruledtabular}
\end{table}

The values of freeze-out parameters such as $T_{ch}$,$\mu_B$, and $\mu_s$ used for this work are in agreement with the THERMUS model fit. A detailed analysis with the THERMUS model can be found in the corresponding papers \cite{Vovchenko:2019pjl,STAR:2017sal}. The exact values of input parameters used for this work can be found in Table~\ref{tab:t3}. 

\begin{table*}
\caption{\label{tab:t3}%
The values of input model parameters for different beam energies.
}
\begin{ruledtabular}
\begin{tabular}{cccccccc}
\textrm{$\sqrt{s_{NN}}$ (GeV)}&
\textrm{$T_{ch}$(GeV)}&
\textrm{$\mu_B$(GeV)}&
\textrm{$\mu_s$(GeV)}&
\textrm{$T_{th}$(GeV)}&
\textrm{$\mu_\pi ^{eff}$(GeV)}&
\textrm{$\eta^{max}$}&
\textrm{$\rho_u^{max}$}\\
\colrule
62.4 & 0.165 & 0.0725 & 0.015 & 0.095 & 0.082 & 2.6 & 0.96 \\
39.0 & 0.164 & 0.112 & 0.028 & 0.092 & 0.088 & 2.1 & 0.94\\
27.0 & 0.1625 & 0.156 & 0.033 & 0.091 & 0.0962 & 1.9 & 0.93\\
19.6 & 0.160 & 0.192 & 0.038 & 0.089 & 0.1 & 1.85 & 0.92\\
11.5 & 0.1526 & 0.268 & 0.055 & 0.085 & 0.1005 & 1.7 & 0.91\\
\end{tabular}
\end{ruledtabular}
\end{table*}
\begin{figure*}

\includegraphics[width=1.07\textwidth]{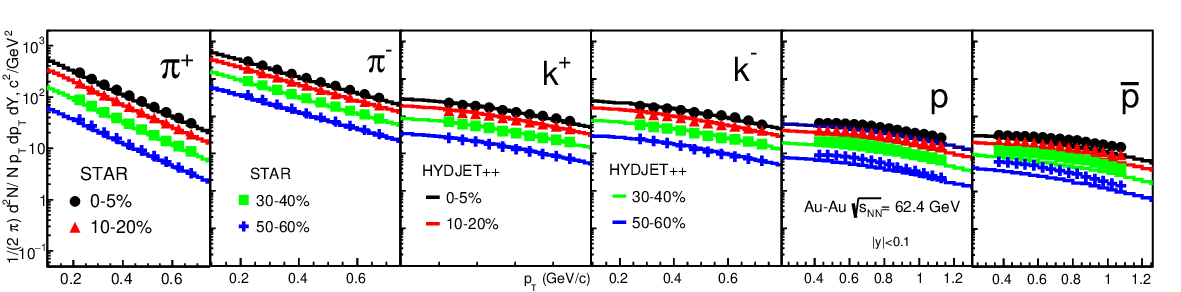}
\caption{\label{fig:5}Transverse momentum spectra for different centralities at $\sqrt{s_{NN}}$= 62.4 GeV. Markers represent experimental data, and HYDJET++ results are shown as lines.}

\end{figure*}

\begin{figure*}
\includegraphics[width=1.07\textwidth]{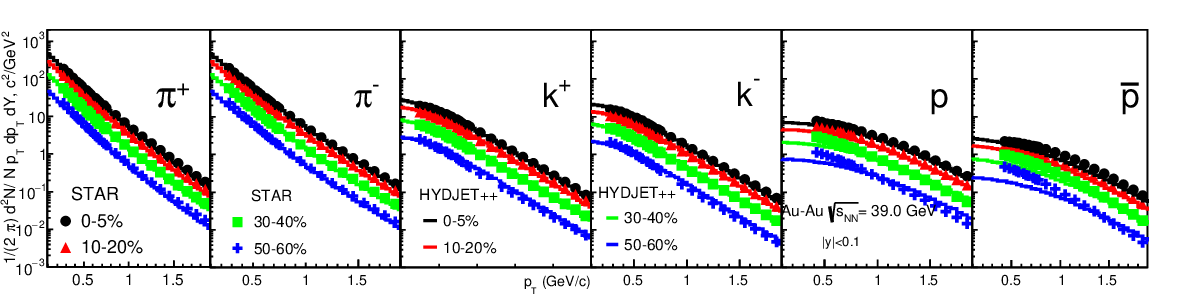}
\caption{\label{fig:6}Transverse momentum spectra for different centralities at $\sqrt{s_{NN}}$= 39.0 GeV. Markers represent experimental data, and HYDJET++ results are shown as lines.}
\end{figure*}
\begin{figure*}
\includegraphics[width=1.07\textwidth]{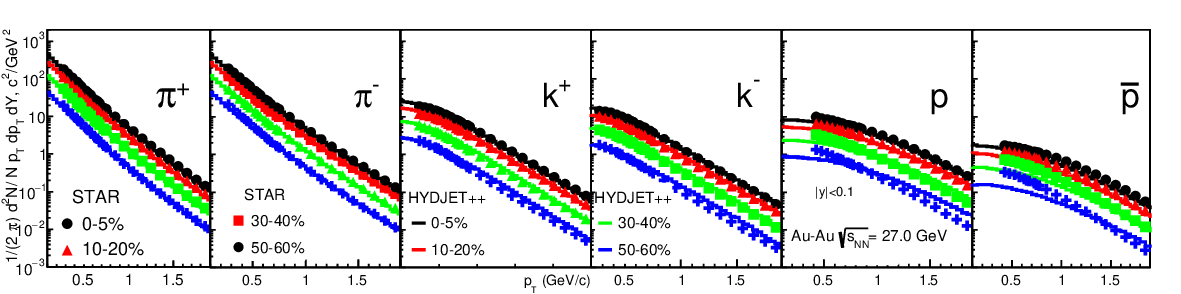}
\caption{\label{fig:7}Transverse momentum spectra for different centralities at $\sqrt{s_{NN}}$= 27.0 GeV. Markers represent experimental data, and HYDJET++ results are shown as lines.}
\end{figure*}

\begin{figure*}
\includegraphics[width=1.07\textwidth]{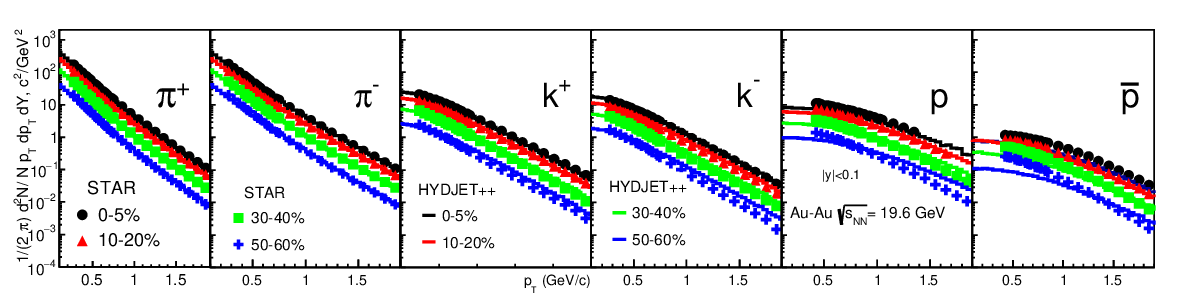}
\caption{\label{fig:8}Transverse momentum spectra for different centralities at $\sqrt{s_{NN}}$= 19.6 GeV. Markers represent experimental data, and HYDJET++ results are shown as lines.}
\end{figure*}
\begin{figure*}
\includegraphics[width=1.07\textwidth]{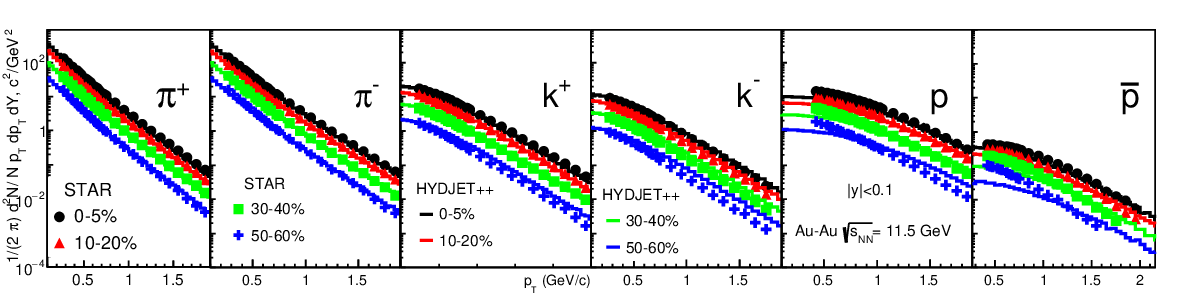}
\caption{\label{fig:9}Transverse momentum spectra for different centralities at $\sqrt{s_{NN}}$= 11.5 GeV. Markers represent experimental data, and HYDJET++ results are shown as lines.}
\end{figure*}

\begin{figure*}
\includegraphics[width=1.07\textwidth]{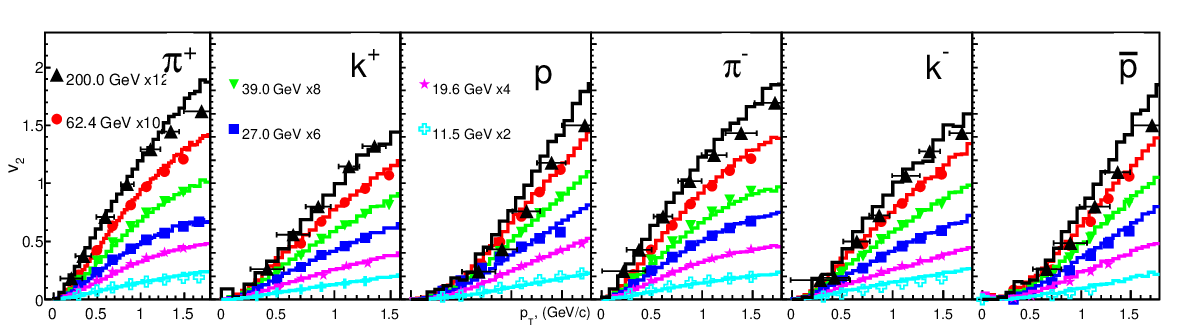}
\caption{\label{fig:10} $v_2$ at different collision energies. Markers represent experimental data. Lines of the same color show HYDJET++ results.}
\end{figure*}
\begin{figure}[h]
\includegraphics[width=.5\textwidth]{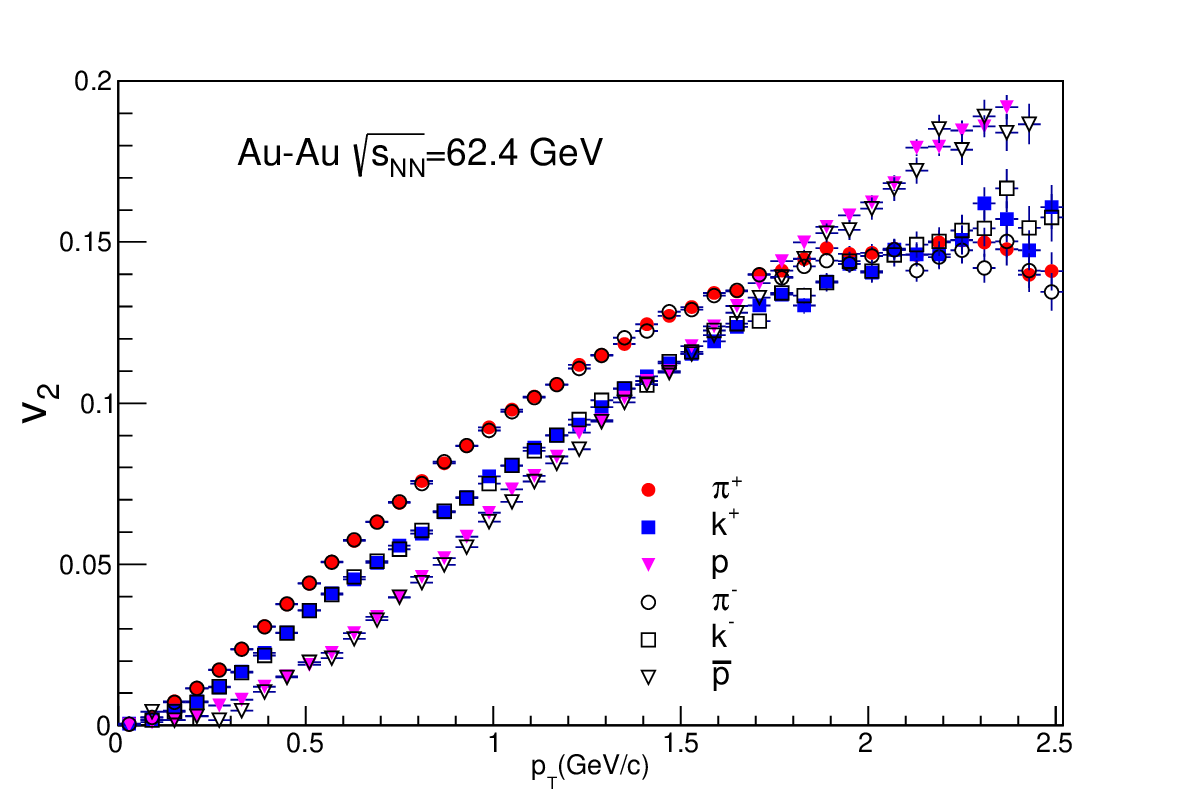}
\caption{\label{fig:12a} $v_2$ of primary hadrons at $\sqrt{s_{NN}}$=62.4 GeV.}
\end{figure}
\begin{figure}[h]
\includegraphics[width=.5\textwidth]{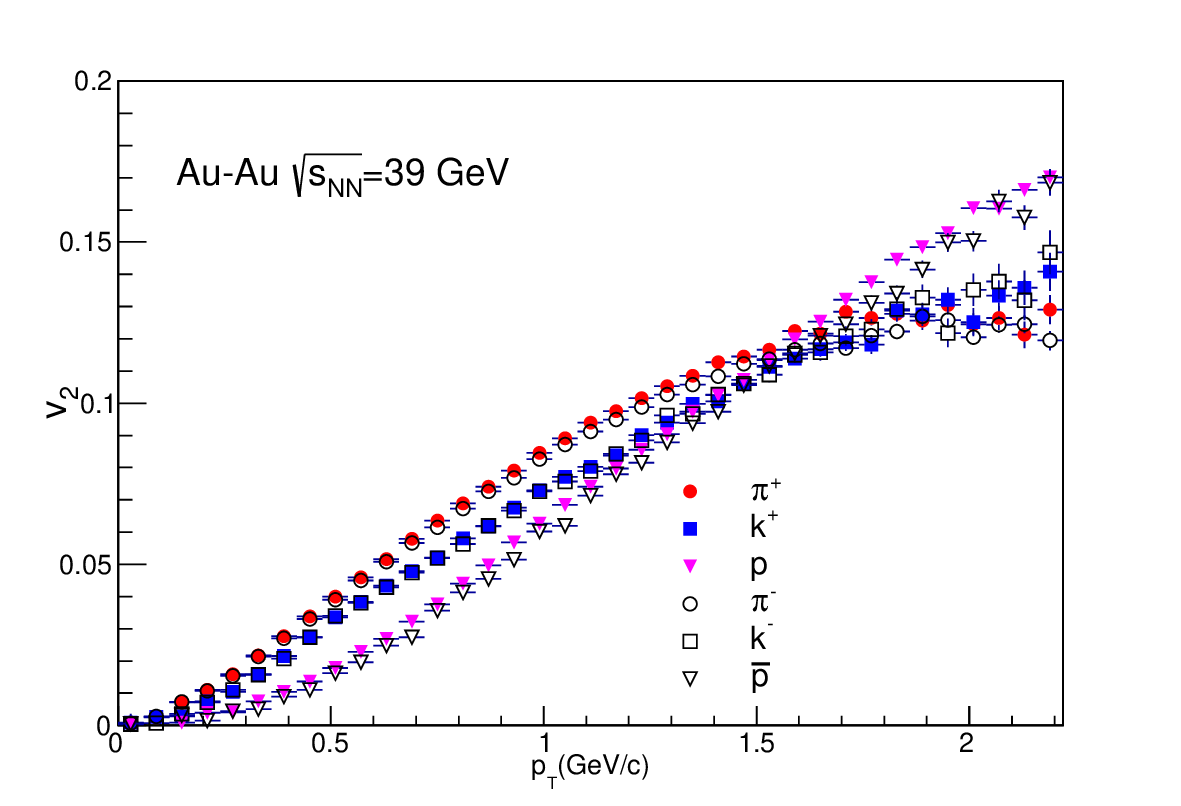}
\caption{\label{fig:12b} $v_2$ of primary hadrons at $\sqrt{s_{NN}}$=39.0 GeV.}
\end{figure}
\begin{figure}[h]
\includegraphics[width=.5\textwidth]{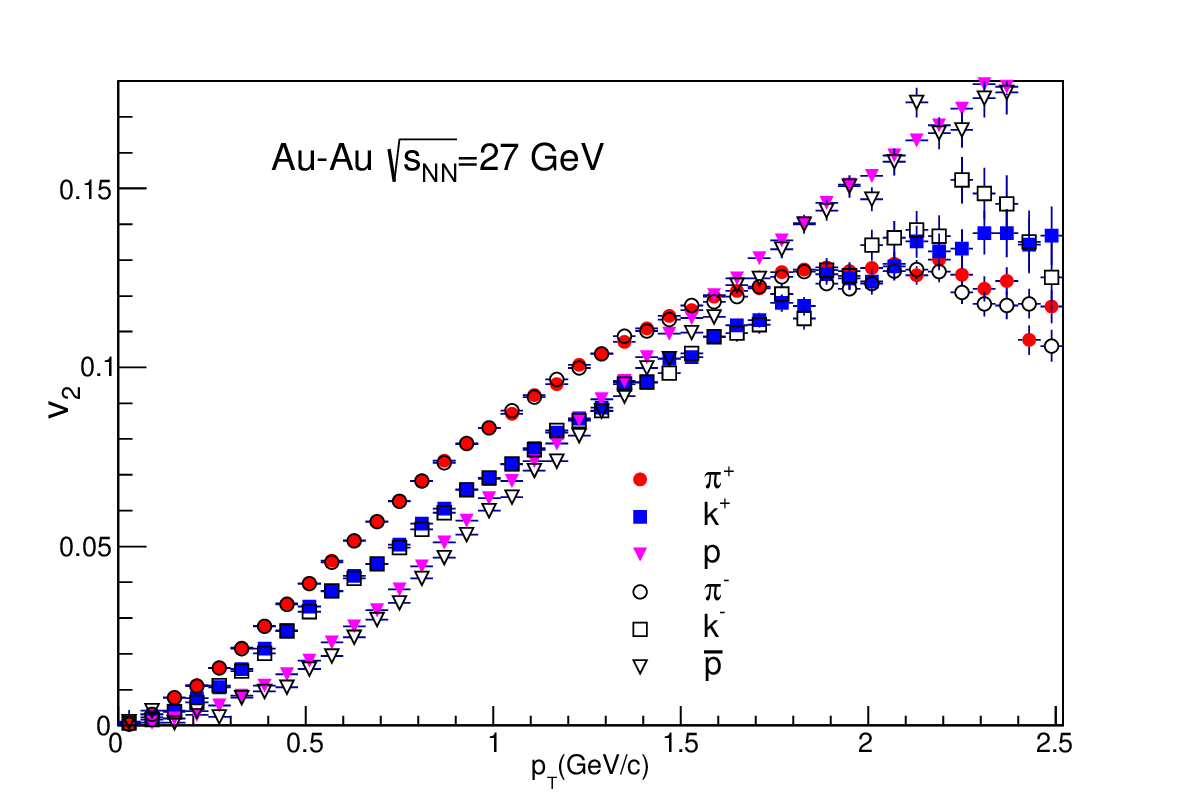}
\caption{\label{fig:12c} $v_2$ of primary hadrons at $\sqrt{s_{NN}}$=27.0 GeV.}
\end{figure}
\begin{figure}[h]
\includegraphics[width=.5\textwidth]{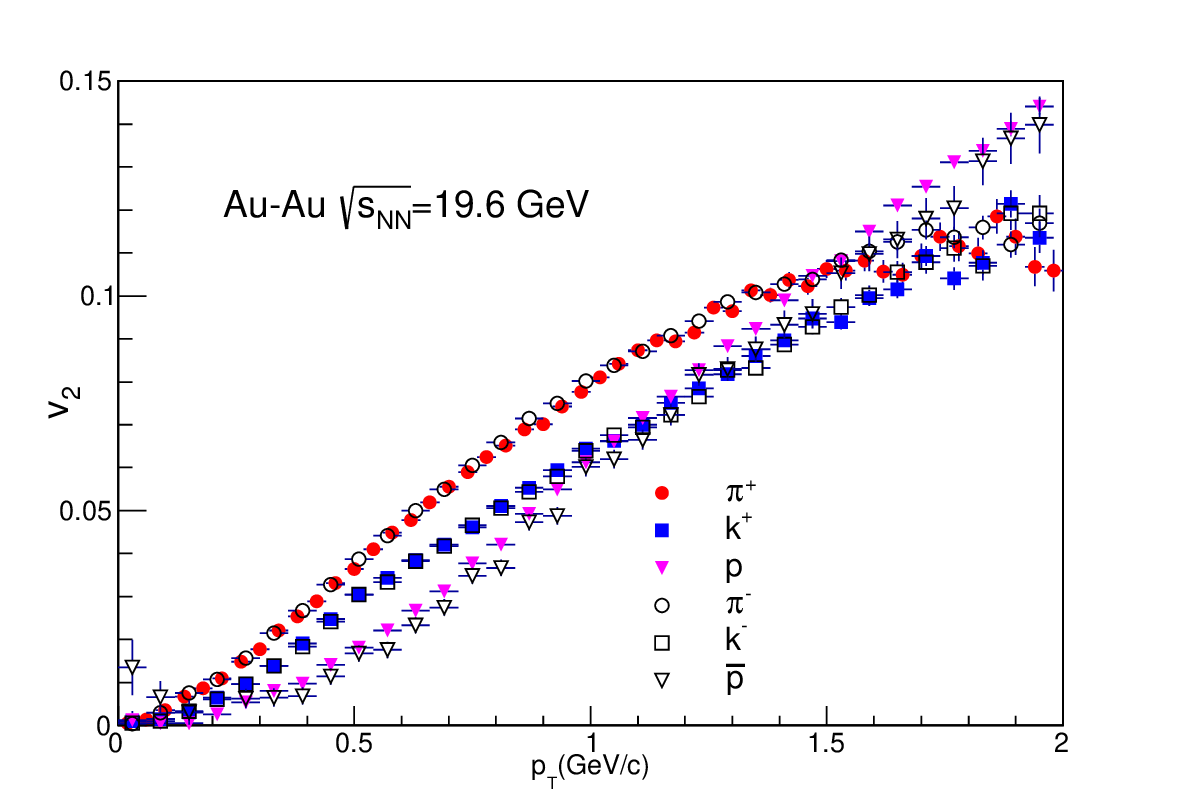}
\caption{\label{fig:12d} $v_2$ of primary hadrons at $\sqrt{s_{NN}}$=19.6 GeV.}
\end{figure}
\begin{figure}[h]
\includegraphics[width=.5\textwidth]{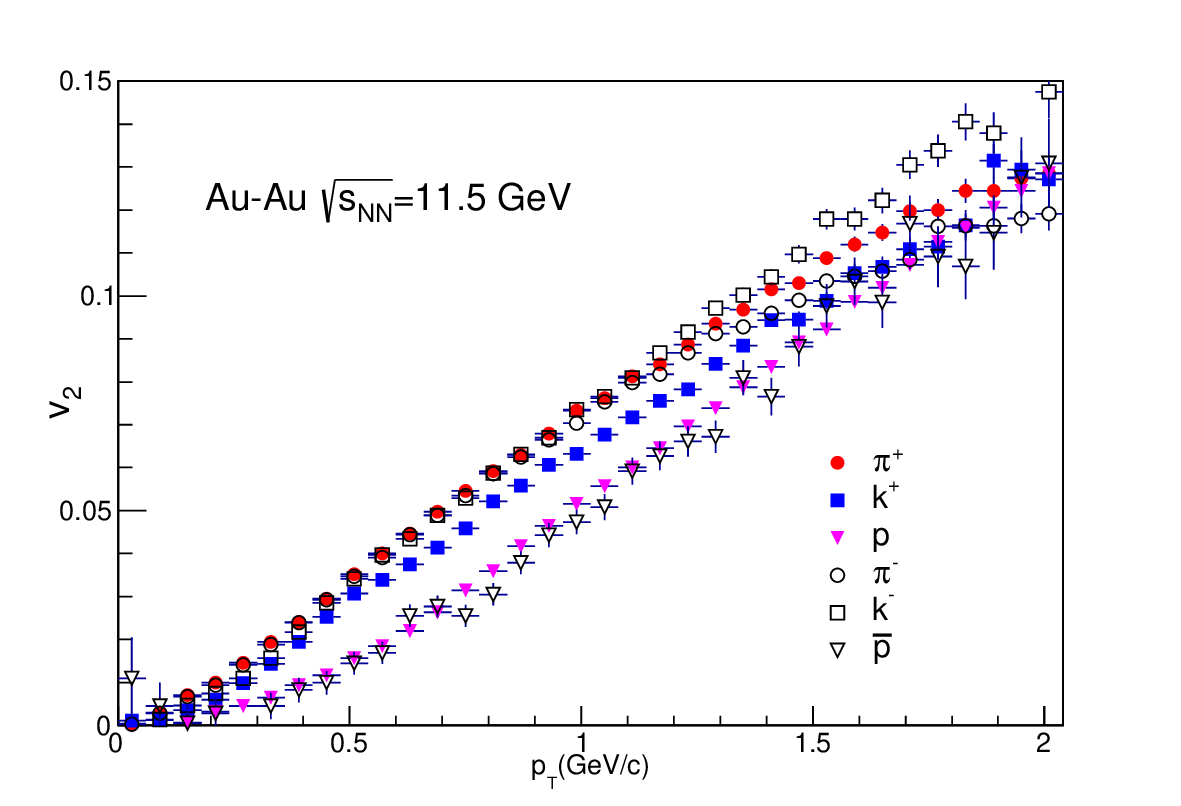}
\caption{\label{fig:12e} $v_2$ of primary hadrons at $\sqrt{s_{NN}}$=11.5 GeV.}
\end{figure}
\subsection{Particle Ratios}
Fig.~\ref{fig:two} shows HYDJET++ results for $\pi^-/\pi^+$,$k^-/k^+$,$k^+/\pi^+$,$k^-/\pi^-$ and $\bar{p}/p$ ratios at various center of mass energies for most central collisions at mid rapidity. The $\pi^-/\pi^+$ ratio increases at lower collision energies but still remains close to 1. The $k^-/k^+$ ratio increases with beam energy and approaches unity at $\sqrt{s_{NN}}=62.4$ GeV. The $\bar{p}/p$ ratio also increases with beam energy, but the increase is more rapid than $k^-/k^+$. The $k^+/\pi^+$ ratios decrease with increasing beam energy, while the $k^-/\pi^-$ ratio shows the opposite trend. Both these ratios become the same after 62.4 GeV due to smaller values of $\mu_s$ at higher collision energies. The model calculations agree well with the experimental data with a slight deviation at 11.5 GeV.

\begin{figure}[h]
\includegraphics[width=.5\textwidth]{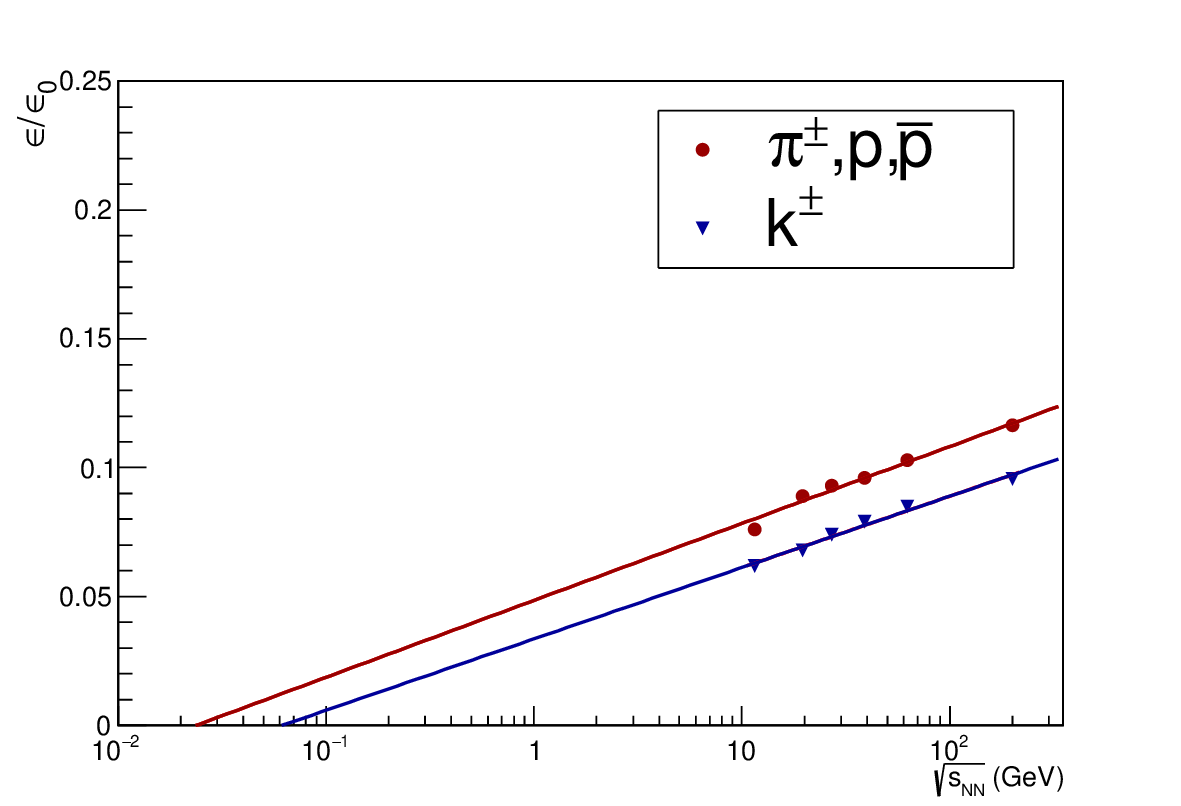}
\caption{\label{fig:11}Relation between scaling factor k and beam energy. The lines show the fitting function from eq. ~(\ref{eq:16}).}
\end{figure}

\begin{figure*}
\includegraphics[width=1.0\textwidth]{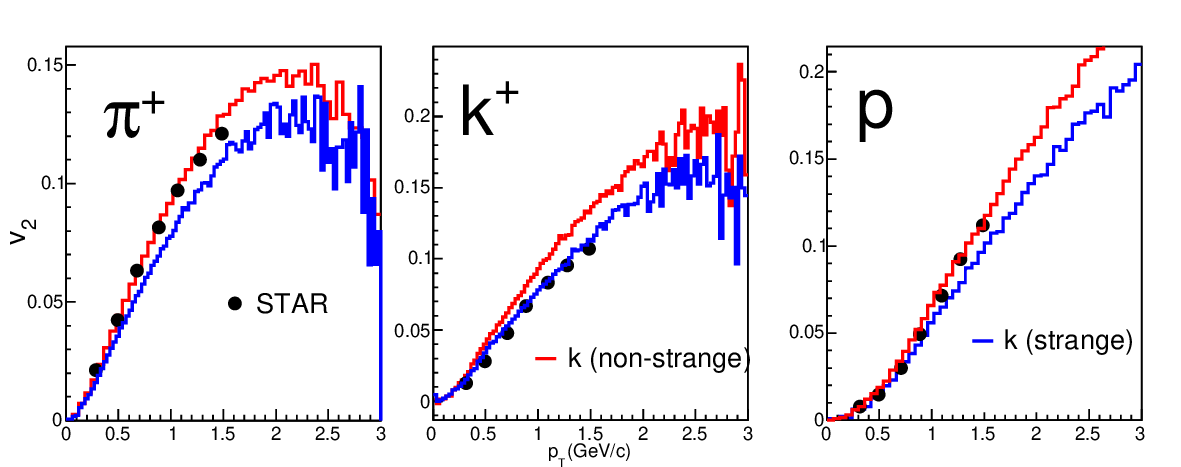}
\caption{\label{fig:12f} $v_2$ of identified hadrons at 62.4 GeV with different $k$ values. The red lines represent the results with the values of $k$ used for non-strange hadrons, and the blue lines represent the results with the values used for strange hadrons. }

\end{figure*}

\subsection{$p_T$ spectra}
In the HYDJET++ framework, the particles are produced as a combination of a hydro part and jets. So, it is crucial to check the contribution from both processes. Fig.~\ref{fig:1} shows the contributions from hydro and jets in $p_T$ spectra of $\pi^\pm$ for most central collisions for $\sqrt{s_{NN}}$=62.4, 27.0 and 11.5 GeV. It is evident that the hydro part predominates the $p_T$ spectra at low $p_T$, and the contribution from jets starts to affect the overall form of $p_T$ spectra at relatively higher $p_T$. It's worth noting that the contributions from both processes drop with decreasing beam energy. However, the contribution from jets decreases more rapidly than that of the hydro part. In Fig. ~\ref{fig:2} we have shown the effect of both processes for $k^\pm$
, the contributions are similar to that of charged pions, however for $k^-$, the contributions from the soft part decrease more rapidly with decreasing beam energy than $k^+$. The effects from hydro and jets for $p$ and $\bar{p}$ are shown in Fig.~\ref{fig:3}. In the $p_T$ spectra of protons, the contributions from jets decrease with decreasing beam energy while the contributions from the hydro part increase rapidly. As a result, the overall contribution increases. This can be attributed to an increase in baryon density at lower collision energies \cite{STAR:2017sal}. For $\bar{p}$, contributions from both processes decrease rapidly with decreasing beam energy. However, the hydro contribution decreases faster than the jets, leading to the modification of low $p_T$ spectra by jets.  We have discussed the difference in $p_T$ of particles and antiparticles later. A similar analysis for Au-Au collisions at 200 GeV using the HYDJET++ model for kaons and pions can be found in the corresponding papers \cite{Singh:2023bzm,Lokhtin:2008xi}.

A clear beam energy dependence is observed in the difference in the invariant yields of $p$ and $\bar{p}$. The difference increases with the decrease in beam energy. This is due to the increase in chemical potential which determines the difference between the number of particles and antiparticles \cite{Anchishkin:2021wxt,Cleymans:1998fq}, the effect of chemical potential is clearly observed in the soft part of the $p_T$ spectra. However, the contribution from jets makes the difference relatively negligible beyond $p_T\approx1.5$ GeV/c. Similar trends are seen for $k^\pm$ as well, but they are less pronounced in comparison to $p$ and $\bar{p}$ because of the smaller values of strangeness chemical potential compared to baryonic chemical potential. Charged pions that have negative chemical potential show an opposite trend at low $p_T$ ($p_T \le 1$ GeV/c ). A detailed comparison can be seen in Fig.~\ref{fig:4}.

In Fig.~\ref{fig:5} to Fig.~\ref{fig:9}, we have shown the transverse momentum spectra at low $p_T$ for different beam energies. Our results are shown in four centrality classes i.e. (0-5)\%, (10-20)\%,  (30-40)\%, and (50-60)\%. For comparison, we have shown the experimental data from STAR\cite{STAR:2017sal,STAR:2008med}. The $p_T$ spectra of $\pi^\pm$ and $k^\pm$ show a good match with the experimental data in central and semi-central collision while the model slightly over-predict $k^\pm$ at peripheral collisions for lower beam energies. The model under-predicts the $p_T$ spectra of $p$ and $\bar{p}$ up to $p_T=0.5$ GeV$/c$, and after 0.5 GeV$/c$ there is a good match. The results for peripheral collisions at lower beam energy show a clear mismatch with the experimental data. This is due to a strong centrality dependence of $\mu_B$, which is observed at 11.5 and 19.6 GeV \cite{STAR:2017sal}. 

The centrality dependence at lower collision energies is similar to the one observed for the top RHIC energy as well as LHC energies \cite{ALICE:2018vuu,PHENIX:2013kod}. The invariant yield tends to decrease from central to peripheral collisions. This behavior is the same for all primary charged hadrons. The ratio of the invariant yields of central and peripheral collision is the same for all collision energies for all particles. Thus, there is no beam energy dependence of variation of $p_T$ spectra with collision centrality observed in our model calculations.

 Recent experiments have shown that high $p_T$ suppression due to jet quenching is not observed for lower collision energies, instead an enhancement is observed, possibly due to the Cronin effect \cite{STAR:2017ieb}. Since the HYDJET++ model does not have any mechanism for enhancement of high $p_T$ hadrons the results will underpredict data. Hence, we have only shown our results at low $p_T$ where no such effects are observed.

\subsection{Elliptic flow ($v_2$)}
In non-central heavy ion collisions, the initial asymmetry of the collision results in a higher pressure gradient along the reaction plane compared to directions out of this plane. One of the major aspects of the particles produced inside the QGP medium of the off-centered collisions is that more particles are emitted along the reaction plane than perpendicular to it \cite{Kharzeev:2000ph,Snellings:2011sz,Poskanzer:1998yz}. Such a phenomenon is called the elliptic flow. The magnitude of elliptic flow is given by 
\begin{eqnarray}
v_2=cos(2\phi)
\label{eq:14}
\end{eqnarray}
where $\phi$ is the spatial azimuthal angle and the reaction plane angle is taken 0.

 The strange quarks can have a lower collectivity than up/down quarks due to their higher mass. The flow of the quarks can't be directly observed in experiments.  To test this aspect, we used the scaling parameter ``$k$" as a proxy for collective behavior.  Hence, in this work, we have treated the strange and non-strange hadrons differently to check the impact of quark content on the scaling factor ``$k$".

In Fig .~\ref{fig:10} we have presented the HYDJET++ model calculations for $v_2$ as a function of $p_T$ for the center of mass energy $\sqrt{s_{NN}}$ = 200.0, 62.4, 39.0, 27.0, 19.6 and 11.5 GeV for minimum bias collisions (0-80\%). For clear representation we have scaled $v_2$ of all energies by a different factor, the exact weights for scaling can be found in the legend. The results are compared with the experimental data from STAR \cite{STAR:2013ayu} and PHENIX \cite{PHENIX:2003qra} experiments. HYDJET++ calculations agree well with the experimental data \cite{STAR:2013ayu,PHENIX:2003qra} up to $p_T$ = 2 GeV/c at mid-rapidity. The model overpredicts the $v_2$ of $p$ and $\bar{p}$ at 200 GeV, it also overpredicts the $v_2$ of $k^-$ at 11.5 GeV. The $v_2$ increases with $p_T$ for all collision energy and reaches a maximum value of approximately 0.15 for mesons and 0.2 for baryons, the maximum value decreases with decreasing beam energy. It should be noted that $v_2$ at low $p_T$ (i.e. $p_T \leq 1$ GeV/c) is independent of collision energy, the effect of collision is observed after 1 GeV/c as $v_2$ approaches its maximum value.

The elliptic flow($v_2$) at lower $p_T$($p_T<1.5$ GeV/c) shows a dependence on the mass of the particle species. The particles with lower mass have a larger value of $v_2$, i.e. ($v_2(\pi^+)>v_2(k^+)>v_2(p)$). A similar trend is also observed for corresponding anti-particles \cite{Huovinen:2001cy,Huovinen:2006jp}. A mass ordering in $v_2$ has been observed in the Au-Au collision at 200 GeV \cite{STAR:2008ftz,STAR:2015rxv}. Recent experiments like RHIC BES have also observed a mass ordering in $v_2$ at lower collision energies \cite{STAR:2013ayu}. To validate our approach of treating strange and non-strange hadrons separately, we should check for mass ordering in our calculations. In Fig.~\ref{fig:12a} to Fig.~\ref{fig:12e}, we have shown $v_2$ as a function of $p_T$ for $\sqrt{s_{NN}}$= 62.4, 39.0, 27.0, 19.6, and 11.5 GeV. The $v_2$ of all the particles are plotted together for a clear comparison. The $v_2$ of particles are represented by solid markers, and the $v_2$ of antiparticles are shown by open markers to get an idea about the difference in $v_2$ of particles and antiparticles, more details can be found in the legend. A clear mass ordering can be seen in the HYDJET++ calculation in all five collision energies. Mass ordering is observed up to $p_T$ = 1.5 GeV/c. After 1.5 GeV/c the $v_2$ of protons increases rapidly and surpasses the $v_2$ of pions and kaons. At about 2 GeV/c an opposite trend is observed i.e ($v_2(\pi^+)<v_2(k^+)<v_2(p)$). For $\pi^\pm$ and $k^\pm$ the $v_2$ is the same for both particles and anti-particles, while at 11.5 GeV $k^-$ has a higher $v_2$ than $k^+$. Protons have a relatively larger $v_2$ than the anti-protons at all collision energies.

We have found that for low $p_T$ ($p_T<1.5$ GeV), the value of the scaling factor ``$k$" in minimum bias collisions has the following relation with beam energy.

\begin{eqnarray}
k=a\ln(\sqrt{s_{NN}})+b
\label{eq:16}
\end{eqnarray}
where, for $\pi^\pm$ and $p\bar{p}$\\
$a=0.0129\pm 0.001$,
$b=0.0485\pm 0.004$,\\
and, for $k^\pm$\\
$a=0.012\pm 0.0007$,
$b=0.0335\pm 0.002$.

The value of $a$ is almost the same for all particles, while the value of $b$ for non-strange hadrons is greater than the strange hadrons. It indicates that in the HYDJET++ framework, the magnitude of azimuthal spatial and momentum anisotropy (which determines the elliptic flow($v_2$)) is dependent on quark content. A lower $k$ for strange hadrons suggests a lower $v_2$ of strange quarks before hadronization. A similar observation was made by J. Song $et$ $al$ \cite{Song:2020uvu}. The study extracted the $v_2$ of quarks based on the Equal Velocity quark Combination (EVC) mechanism and found that strange quarks have smaller $v_2$ compared to up/down quarks. However, more studies are required to know the exact reason for such behavior. The collective behavior is weaker for the lower energies of the Beam Energy Scan range. Hence, the particles are emitted uniformly in all directions \cite{Slodkowski:2021cld}, which reduces their spatial and momentum anisotropy, which reflects as a reduced ``$k$" value \cite{Nayak:2024iio}. A clear depiction can be found in Fig.~\ref{fig:11}. 

In Fig.~\ref{fig:12f}, we have shown the elliptic flow of $k^+$, $\pi^+$, and p in minimum bias Au-Au collisions at 62.4 GeV with different values of ``$k$". It can be seen that the HYDJET++ calculations are sensitive to ``$k$", a small change in ``$k$" can have a notable change in $v_2$. The effect is more distinguished for the mesons than protons. If the value of ``$k$", which was used for non-strange hadrons is used for strange hadrons, it would significantly overpredict the data, and vice versa. Hence, different values of $k$ should be used for both cases to better describe the data.

\section{Summary}
We have performed a comprehensive study of bulk properties of medium like $p_T$ spectra and elliptic flow produced in Au-Au collisions at different beam energies. HYDJET++ model shows a good match with the experimentally observed particle ratios, which validates the Cleymans-Reidlich parameterization of $T_{ch}$ and $\mu_B$ for lower collision energies. HYDJET++ calculations of $p_T$ spectra are in agreement with experimental data for central and semi-central collisions. The lower energies of the BES have a higher chemical potential and a lower freeze-out, these conditions affect the $p_T$ distribution of p and $\bar{p}$. A large $\mu_B$ favors the thermal proton production while reducing the anti-proton production. The jet contribution also reduces due to a decrease in inelastic cross-section ($\sigma_{NN}^{inel}$) with beam energy. The interplay between these effects can be seen in the $p_T$ spectra of $\bar{p}$ where the jet contributions affect the low $p_T$ spectra. The model fails to describe the p and $\bar{p}$ spectra in peripheral collisions, this can be attributed to the centrality dependence of $\mu_B$, which is observed for these collision energies.

We have re-interpreted elliptic flow ($v_2$) based on the scaling of azimuthal spatial anisotropy. This approach allows us to calculate $v_2$ without having to track the whole hydrodynamic evolution of the system. The scaling between initial and final spatial anisotropy, referred to as ``$k$" in this work, is used to calculate $v_2$. The approach can successfully describe the experimental data up to 11.5 GeV. It also preserves the essential aspects of $v_2$ such as mass ordering at both low $p_T$ ($v_2(\pi^+)>v_2(k^+)>v_2(p)$) and high $p_T$ ($v_2(p)>v_2(\bar{p})$). The value of ``$k$" is found to be a function of beam energy and can be parameterized as a function of $\ln{\sqrt{s_{NN}}}$. The primary mesons are found to be more sensitive to $k$ than the baryons. A different value of ``$k$" is required for strange and non-strange hadrons to better describe the data, which suggests that the relatively heavier strange quarks have smaller spatial and momentum anisotropy. However, this observation is empirical in nature, and more detailed studies are needed to have a better understanding of the transport properties of strange quarks.

\begin{acknowledgments}
BKS sincerely acknowledges financial support from the Institute of Eminence (IoE) BHU Grant number 6031. SRN acknowledges the financial support from the UGC Non-NET fellowship during the research work. SP
acknowledges the financial support obtained from UGC and IoE under a
research fellowship scheme during the work.

\end{acknowledgments}
\section*{Data availabilty}
 The data will be made available on request.

\end{document}